\documentclass[printer]{aa}
\usepackage{amssymb}
\usepackage{amsmath}
\usepackage{txfonts}
\usepackage{graphicx}
\usepackage{natbib}
\usepackage{rotating}
\usepackage{soul}
\usepackage{color}
\usepackage{url}
\usepackage{epsfig}
\usepackage{longtable}
\usepackage{ulem}

\usepackage{color}

\begin{document}

\title{Neutrino signal from extended Galactic sources in IceCube}
\titlerunning{Galactic sources in IceCube}

\author{ C. Tchernin$^1$,  J.A. Aguilar$^2$, A. Neronov$^1$, T. Montaruli$^2$}
\institute{ 1. D\'epartement d'astronomie, Universit\'e de Gen\`eve, CH-1290 Versoix, Switzerland.\\
2. D\'epartement de physique nucl\'eaire et corpusculaire,
Universit\'e de Gen\`eve, CH-1211 Gen\`eve 4, Switzerland.}


\abstract
{The Galactic Plane is the brightest source of gamma rays in the sky. It should be also (one of the) brightest very-high-energy neutrino sources, if a neutrino flux comparable to the $\gamma$-ray flux is  produced by the cosmic ray interactions in the interstellar medium. }
{ We explore the detectability of the neutrino flux from the entire Galactic Plane or from a part of it with IceCube. }
{We calculate the normalization and the spectral index of the neutrino power law spectrum from different regions of the Galactic plane, based on the observed spectral characteristics of the pion decay $\gamma$-ray diffuse emission observed by the Fermi/LAT telescope in the energy band above 100~GeV. We compare the neutrino flux calculated in this way with the sensitivity of IceCube for the detection of extended sources. }
{Assuming a binned extended source analysis method, we find that the only possible evidence for neutrino emission for sources located in the Northern hemisphere is from the Cygnus region after 20 years of exposure. For other parts of the Galactic Plane even a 20 year exposure with IceCube is not sufficient for the detection. Taking into account marginal significance of the detectable source in the Cygnus region, we find a precise position and size of the source region which optimizes the signal-to-noise ratio for neutrinos. We also calculate the low-energy threshold above which the neutrino signal could be detected with the highest signal-to-noise ratio. This calculation of precise source position, size and energy range, based on the gamma-ray data, could be used to remove the "trial factor" in the analysis of the real neutrino data of IceCube.  We notice that the diffuse neutrino emission from the inner Galactic Plane in the Southern Hemisphere is much brighter. A neutrino detector with characteristics equivalent to IceCube, but placed at the Northern Hemisphere (such as KM3NeT), would detect several isolated neutrino sources in the Galactic Plane within just five year exposure at 5$\sigma$ level. These isolated sources of $\sim$TeV neutrinos would unambiguously localize sources of cosmic rays which operated over the last 10 thousand years in the Galaxy.}
{The detection of the diffuse neutrino emission from cosmic ray interactions in the Galactic Disk is challenging, but marginally feasible with 20 yr long exposure by IceCube.  Dramatically shorter (2-5~yr) exposure times would be required for the Galactic Plane signal detection with an IceCube like neutrino detector in the Northern hemisphere.}  
\maketitle

\section{Introduction}

The Galactic Plane of our Milky Way galaxy is the brightest source of $\gamma$-rays \citep{oso-3,sas-2,cos-b2,egret,Fermi_sky,HESS_survey}. The  $\gamma$-ray emission from the Galaxy originates from the cosmic ray interactions in the interstellar medium. These interactions result in production of neutral and charged pions which subsequently decay into $\gamma$-rays, electrons/positrons and neutrinos.  The overall power and spectral characteristics of the $\gamma$-ray and neutrino emission produced by the broad-band distribution of cosmic rays are very similar to each other \citep{stecker79,kamae,kelner}. 

The $\gamma$-ray emission from the decays of neutral pions is directly  observed in the 0.1-100~GeV band by the Large Area Telescope (LAT) on board of the Fermi satellite  \citep{Fermi_sky}. At the higher energies, the ground based $\gamma$-ray telescopes, HESS, MAGIC  and VERITAS, have recently started to explore the Galactic Plane. 
HESS survey of the inner part of the Galactic Plane has revealed a number of bright extended $\gamma$-ray sources \citep{HESS_survey,HESS_survey1}, but no continuous diffuse emission from the entire Galactic Plane, except for the Galactic Ridge region \citep{Gal_ridge}. The same bright extended sources are also seen in the survey of the Galactic Plane above 100~GeV performed by Fermi/LAT \citep{neronov_semikoz}. The appearance of isolated extended sources in the Galactic plane above 100 GeV, rather than of a continuous emission due to synchrotron and inverse Compton of leptons, could be explained if cosmic rays take about 10$\%$ of the average energy of $\sim 10^{51}$ erg of an average number of one supernova explosion per century. In this case, the shorter escape time of cosmic rays with energies above $\sim 1$~TeV (compared to lower energy cosmic rays) from the Galactic Disk does not allow them to spread homogeneously throughout the Disk \citep{felix_kniga,neronov_semikoz}. Locating the extended sources of pion decay $\gamma$-ray emission in the $E>100$~GeV band one gets a possibility to identify all the places in which injection of cosmic rays happened over the last $10^4-10^5$~yr. It is possible that the extended sources in the Galactic Plane revealed in the $E>100$~GeV band by Fermi/LAT and by the ground-based $\gamma$-ray telescopes are in fact the places of recent injections of $E>1$~TeV cosmic rays in the Galaxy \citep{neronov_semikoz}. Verification of this hypothesis implies establishment of  the "hadronic" nature of these sources. The most straightforward way to verify if the observed $\gamma$-ray emission from these sources is produced by the cosmic ray interactions would be to detect a very-high-energy (VHE) neutrino flux from these sources.    

Up to recently, the sensitivity of detectors of VHE neutrinos (AMANDA, ANTARES) was not sufficient for the detection of neutrinos from astronomical sources, including the extended sources in the Galactic Plane. The situation has changed with the completion of the IceCube telescope construction~\citep{icecube}. The IceCube sensitivity to neutrino fluxes is comparable to that of the existing space and ground-based $\gamma$-ray telescopes, although in a somewhat higher energy band (TeV-PeV, rather than GeV-TeV). As shown in another paper \citep{tchernin} the sensitivity of the IceCube is comparable to the observed gamma fluxes.

The unambiguous relation between the $\gamma$-ray and neutrino emission from the cosmic ray interactions in the interstellar medium of the Milky Way makes it possible to obtain a reliable estimate of the flux and spectrum of the neutrino signal from the Galactic Plane \citep{stecker79}. Information on the $\gamma$-ray emission from the Galactic Plane is available up to the $\sim 100$~GeV energy band, i.e. at about the energy threshold of IceCube. 
This means that the expected neutrino signal from the Galactic Plane in IceCube  can be predicted rather accurately. We investigate if this signal from the Galactic Plane is detectable within a multi-year exposure of IceCube.

Previous studies of the detectability of the diffuse neutrino signal from the Galactic Plane based on SAS-2 \citep{stecker79}, COS-B \citep{berezinsky93} and EGRET $\gamma$-ray data at lower energies \citep{evoli07} suffered from a large uncertainty of extrapolation of the $\gamma$-ray data to the multi-TeV energy band in which VHE neutrino telescope(s) are most sensitive. A general conclusion of the early estimates \citep{stecker79,berezinsky93,evoli07} is that the detection of the diffuse neutrino emission from the Galaxy is a challenging task even for the km$^3$ type neutrino detector.  

A detailed calculation of the neutrino flux from the Cygnus region was carried out based on the MILAGRO $\gamma$-ray data in the multi-TeV band \citep{montaruli07,beacom07,taylor08}. The common  conclusions of  these works is that the Cygnus region is detectable within a decade of data taking by IceCube.

 For example, \citet{halzen08}  predicted the extented neutrino emission using the Milagro source fluxes detected with average energy of 3 TeV. Three of these sources are in the Cygnus region. A stacked signal from these sources could be detected if, as in \citet{halzen08}, we assume a spectral index of 2.1 and energy cut-off  of about 600 TeV. Recent gamma-ray observations reported much lower energy cut-offs for these sources \cite{Milagro_sources}, while also not confirming one of the candidates as a real source (significance  $<3\sigma$). One should note, however, that the MILAGRO sources could well be just the brighter spots of the extended emission in the entire Cygnus region (as suggested by the Fermi data, see below). In this respect, the spectra of the brighter TeV spots are not representative for the overall emission spectrum of the entire region.


Limits on the neutrino emission from point sources in the Cygnus region have been set by \citet{icecube}. In addition, a dedicated scan for point sources in the Galactic Plane region was also performed by the IceCube collaboration~\citep{icecube_galplane} as well as search optimized for multiple and extended sources in the Cygnus region. This search was based on the data of the 22-string and 40-string configurations of IceCube together with the AMANDA data. In both cases no evidence of neutrino signal was found but the collaboration set the most stringent 90$\%$ confidence level muon neutrino flux upper limits in the visible Northern Sky for soft-spectra point sources to be in the range $E^{3}dN/dE\sim5.4-19.5\times10^{-11}$TeV$^{2}$cm$^{-2}$s$^{-1}$. 
Moreover, limits are reported for the Cygnus region of about 0.7 (5) $\times10^{-11}$ TeV$^{-1}$ cm$^{-2} $s$^{-1}$ (E/TeV)$^{-\gamma}$ for E$^{-\gamma}$ and $\gamma$ = 2 (3).

In what follows we investigate the perspectives for the detection of a neutrino signal from the Galactic Plane with IceCube, based on the measurement of the $\pi^0$ decay gamma-ray signal by Fermi/LAT. We calculate the  sensitivity of IceCube for extended sources with different spectral characteristics generalizing a previously developed method \citep{neronov09} tested on the published IceCube data for point sources \citep{tchernin}. We then apply this method to the IceCube detector and study which part of the Galactic Plane located in the Northern hemisphere could be detected within a given exposure time. Taking into account that the brightest part of the Galaxy, including the Galactic Center region, is located in the Southern Hemisphere, we study also the detectability of the neutrino signal from the Galaxy with a detector which has characteristics equivalent to IceCube, but situated at the North Pole. We find that such a detector would have a rich astronomical observations program, with some $\sim 3$~extended Galactic sources detectable within a relatively short (five year) exposure. This provides a strong argument in favor of the deployment of the km$^3$ scale VHE neutrino detector in the Northern hemisphere, like KM3NeT  \citep{km3net}.     

\section{The VHE neutrino signal from the Galactic Plane}

The VHE neutrino signal from the Galactic Plane could be reliably estimated based on the observed $\gamma$-ray signal in the GeV energy band, which lies below the energy band in which IceCube is most sensible. A complete picture of the $\gamma$-ray emission from the Galactic Plane in the energy band above 100~GeV is provided by the Fermi/LAT observations, which cover the entire Galactic Plane with approximately homogeneous exposure of $\simeq 4$~yr. 

\begin{figure}
\includegraphics[width=\columnwidth,angle=0]{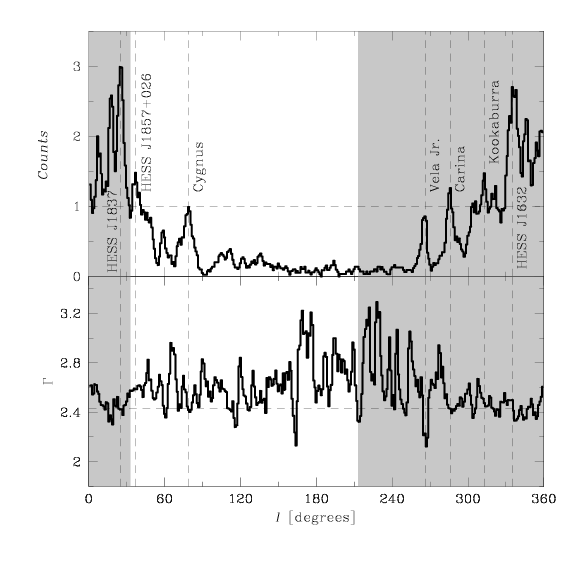}
\caption{Profiles of the intensity (top) and the spectral index (bottom) of the $\gamma$-ray power law emission in the energy band above $100$~GeV. Fermi/LAT counts are collected from the circular regions of radius $\Psi=2^\circ$ around each Galactic longitude of the Galactic plane. The fluxes are normalized on the flux from the Cygnus region at longitude $l=80^\circ$. The shaded region marks the part of the Galactic Plane in the Southern hemisphere.  Vertical lines with the names show locations of bright extended VHE $\gamma$-ray sources.}
\label{fig:profile_gamma}
\end{figure}

\begin{figure}
\includegraphics[width=\columnwidth,angle=0]{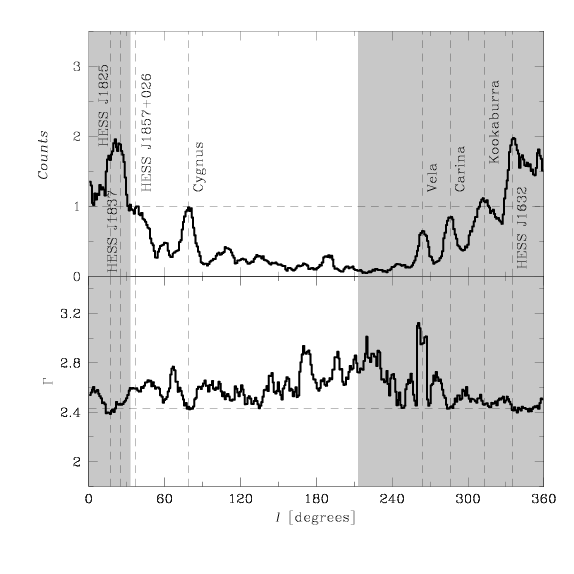}
\caption{Same as in Fig. \ref{fig:profile_gamma}, but for the source region radius $\Psi=4^\circ$.}
\label{fig:spectrum_gamma}
\end{figure}

Fig. \ref{fig:profile_gamma} and \ref{fig:spectrum_gamma} show the profiles of the intensity of the emission from the Galactic Plane as a function of Galactic longitude for circular regions of radii $ 2^\circ$ and $ 4^\circ$ around each Galactic longitude of the Galactic plane. In the energy band above 10~GeV the signal from the Galactic Plane is dominated by the diffuse emission, with only the minor contribution from the point sources (e.g. pulsars, which have high-energy cut-offs in the $E\le 10$~GeV range).  To produce these figures we have used the LAT data collected over the period from August 2008 till October 2012. We have filtered the LAT events using {\it gtselect, gtmktime} tools, following the recommendations of the LAT collaboration\footnote{\tt http://fermi.gsfc.nasa.gov/ssc/data/analysis/}, and retained only events belonging to the "clean" class ({\tt evclass=3}) which are most likely gamma-rays. We have estimated the flux by dividing counts by exposure calculated using the {\it gtexposure} tool. 

From this figure one could see that the brightest emission from the Galactic Plane in the Northern hemisphere comes from two equally bright locations: a region at approximately zero declination (Galactic longitude $\simeq 33^\circ$, around HESS J1857+028), and  from the Cygnus region at Galactic longitude $\simeq 80^\circ$. 

This indicates that the Galactic Plane neutrino signal in IceCube will appear as isolated extended excesses at the positions of the Cygnus region and/or $l=33^\circ$. If one aims at the initial discovery of the neutrino emission from the Galaxy, it appears reasonable to concentrate on the detection of the emission from the Cygnus region and/or $l=33^\circ$. Considering a larger part of the Galactic Plane would just dilute the excess from these two regions through the addition of a stronger background (generated by atmospheric neutrinos) without a significant increase of the signal.  

The calculation of the neutrino signal in IceCube from the $\gamma$-ray data requires the knowledge of not only the normalization $A_{100}$ of the neutrino flux at $100$~GeV, but also of the spectral index, $\Gamma$, of the neutrino spectrum.
\begin{equation}
\label{eq:spectrum}
\frac{dN_\nu}{dE}=A_{100}\left(\frac{E}{100\mbox{ GeV}}\right)^{-\Gamma}.
\end{equation}
To convert the normalization of the gamma-ray signal to the muonic neutrino signal, we use the relation expressed in \cite{kappes07}. As the spectral index for neutrinos from pion decays is nearly identical to that of the $\gamma$-rays, we can derive it directly from the $\gamma$-ray data. Bottom panels of Figs. \ref{fig:profile_gamma} and \ref{fig:spectrum_gamma}  show the spectral index of the neutrino / $\gamma$-ray spectrum derived from the ratio of the integrated fluxes $N_{10-31.6}$, $N_{31.6-100}$ in the $10-31.6$~GeV and $31.6-100$~GeV energy ranges (i.e. from the "hardness ratio" between the two bands):
\begin{equation}\label{eq:Gamma}
\Gamma=1-2\log\left(N_{31.6-100}/N_{10-31.6}\right)
\end{equation}
It is clear that a harder spectrum (lower $\Gamma$ values) implies a higher neutrino flux in the energy range above $\sim 1$~TeV, in which the IceCube detector is most sensitive. From Fig.s~\ref{fig:profile_gamma} and~\ref{fig:spectrum_gamma} one can see that of the two strongest 100~GeV excesses, the Cygnus region spectrum is harder than that corresponding to the excess located at $l=33^{\circ}$. Therefore the neutrino signal in IceCube from the direction of Cygnus should be significantly stronger than that from the $l=33^\circ$ region.

Thus, the Cygnus region should be considered as the best candidate for the possible IceCube discovery of the neutrino signal from the cosmic ray interactions in the Galactic disk. Taking this into account, an additional optimization of the parameters of the IceCube observation of the Cygnus region, based on the $\gamma$-ray data, appears reasonable. 

\subsection{The neutrino signal from the Cygnus region}

\begin{figure}
\includegraphics[width=\linewidth]{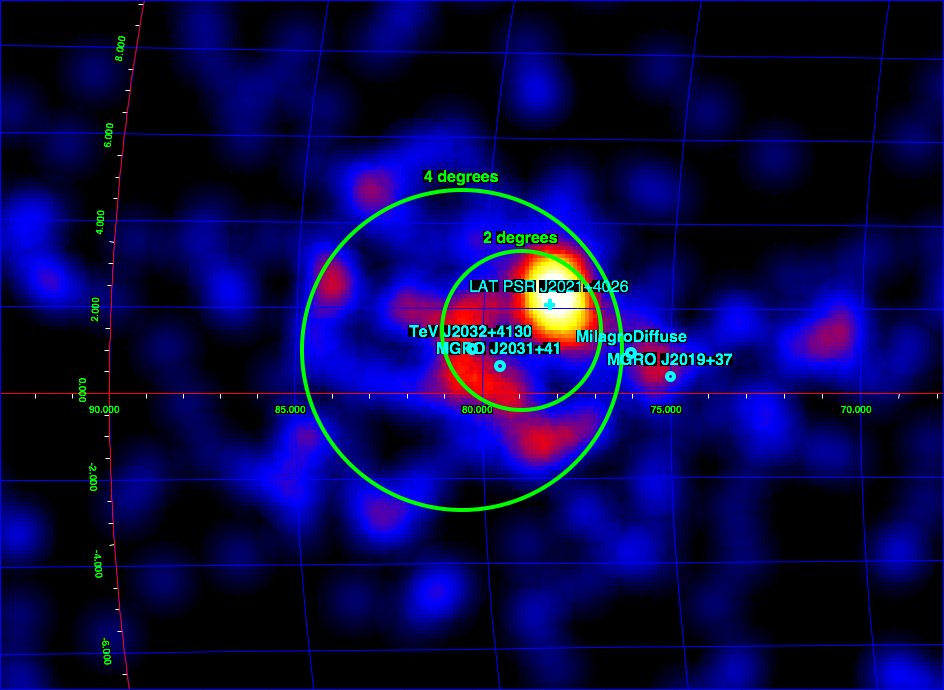}
\caption{The Fermi/LAT countmap of the Cygnus region in the energy band above 100 GeV, smoothed with a Gaussian kernel of 1 degree. The regions indicated by the green circles are used for the spectral extraction of Fermi/LAT data and have radii 2 and 4 degrees (from small to large). Positions and names on known VHE $\gamma$-ray sources are marked. }
\label{fig:cygnus}
\end{figure}

Fig. \ref{fig:cygnus} shows Fermi/LAT countmap of the Cygnus region in the energy band above 100~GeV, smoothed with a 1 degree Gaussian to highlight the extended structures. The dominant source in this energy band is the $\gamma$-Cygni supernova remnant, which contains a pulsar wind nebula and a shell-type supernova remnant. The overall extent of the remnant is about $1^\circ$. The source is detected both in the space-based telescope Fermi/LAT \citep{gammacygni_lat} and ground-based telescope VERITAS \citep{gammacygni_veritas}. At the energies below $\sim 10$~GeV, the emission from the source is strongly dominated by the pulsed magnetospheric emission from the pulsar PSR J2021+4026 \citep{psrj2021}. Above 10~GeV the pulsed emission is strongly suppressed and a separate powerlaw type emission component is present. 

\begin{figure}
\includegraphics[width=\columnwidth]{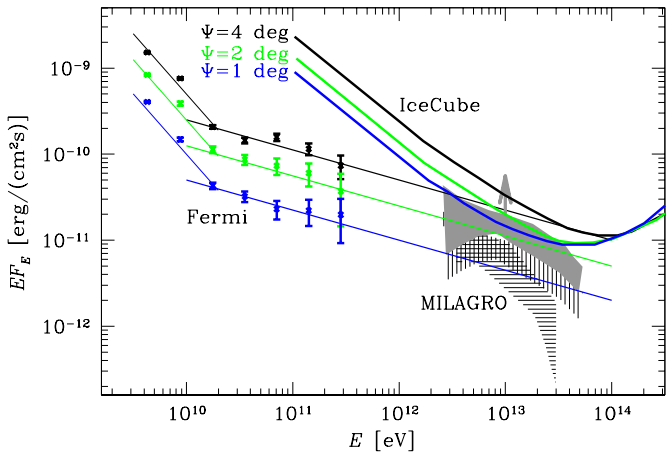}
\caption{Data points: Fermi/LAT spectra of the 1,2 and 4 degrees regions in the Cygnus region (from bottom to top). Straight lines show fits to the spectra with extrapolations to higher energies. Curves represent the 90\% C.L. sensitivities of IceCube (2 yr exposure) for the extended sources of the size 1,2 and 4 degrees (from bottom to top) for all neutrino flavors, using that at Earth, after oscillations, the muonic neutrino flux is one third of the total neutrino flux  \citep{oscillations}. The horizontal hatched region shows the spectrum of the VHE $\gamma$-ray source MGRO J2019+37. The vertical hatched region shows the spectrum of MGRO J2031+41. The grey shading shows the sum of the fluxes of MGRO J2019+37 and MGRO J2031+41 \citep{Milagro_sources}, both sources are within the 4 degree source region considered for spectral extraction in Fermi/LAT. }
\label{fig:cygnus_spectrum}
\end{figure}

The spectrum of the $\gamma$-ray emission from a circular region of radius $\Psi=1^\circ$ centered on the $\gamma$-Cygni supernova remnant position is shown in Fig. \ref{fig:cygnus_spectrum}. To extract the spectrum, we used the "aperture photometry" method. We estimated the diffuse background at the source position from  two regions at the same Galactic latitude, but displaced by five degrees from the source region.  From Fig. \ref{fig:cygnus_spectrum} one could clearly see the presence of the new powerlaw type component in the spectrum above 10~GeV. The flux from this component is detected at least up to 300~GeV in Fermi/LAT. Taking into account the detection of the source by VERITAS, one could conclude that the power law component extends up to the TeV energy band. 
  
The same powerlaw component is present also on a larger angular scale. Fig. \ref{fig:cygnus_spectrum} demonstrates that the flux of this component accumulates on the angular scales $\Psi=2^\circ$ and further on the scale of $\Psi=4^\circ$.  The spectral index of the powerlaw, $\Gamma\simeq 2.4$, remains remarkably stable on different angular scales. This powerlaw component was previously noticed by \citet{gammacygni_lat}, who interpreted this hard component as being produced by the "freshly accelerated" cosmic rays in the Cygnus region. Such an interpretation automatically implies the presence of the hard spectrum neutrino emission from this region, with flux and spectral index nearly identical to the $\gamma$-ray flux. 

No sign of high-energy cut-off of this component is found in the Fermi data. Moreover, observations of the region in the multi-TeV energy band by VERITAS \citep{cygnus_veritas} and Milagro \citep{cygnus_milagro} reveal a strong emission at the energies up to ten(s) of TeV. The measurement of the flux of multiple sources in the Cygnus region seen by MILAGRO in the 10~TeV band agrees well with a simple powerlaw extrapolation of the Fermi/LAT spectrum of extended emission from the Cygnus region. This agreement allows us to make a conjecture that the emission found by MILAGRO is just the high-energy counterpart of the cosmic ray powered $\gamma$-ray emission from the  freshly accelerated cosmic rays injected either by the Cygnus OB2 association \citep{gammacygni_lat} or by the $\gamma$-Cygni / PSR J2021+4026 composite supernova remnant \citep{neronov_semikoz}.

\begin{figure}
\includegraphics[height=\columnwidth,angle=-90]{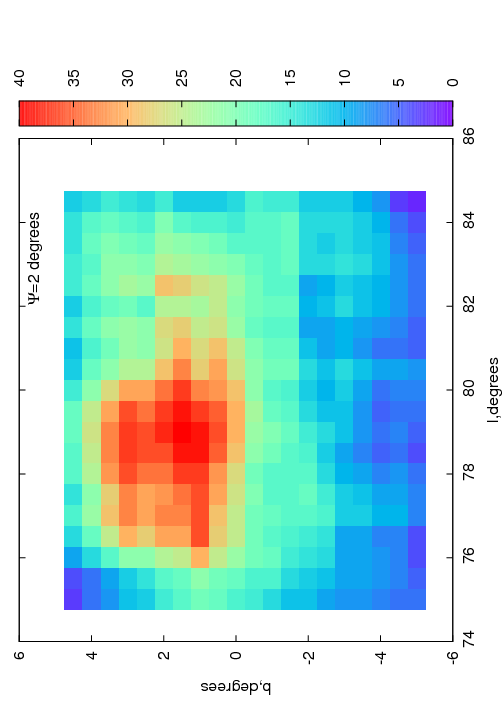}
\caption{The strength of  the $\gamma$-ray (and hence neutrino) signal within circles of radius $\Psi=2^\circ$ at different locations inside the Cygnus region. Each pixel shows the position of the center of the circle.  The color scale shows the number of gamma-ray counts in the circles. }
\label{fig:best_circle_2}
\end{figure}

\begin{figure}
\includegraphics[height=\columnwidth,angle=-90]{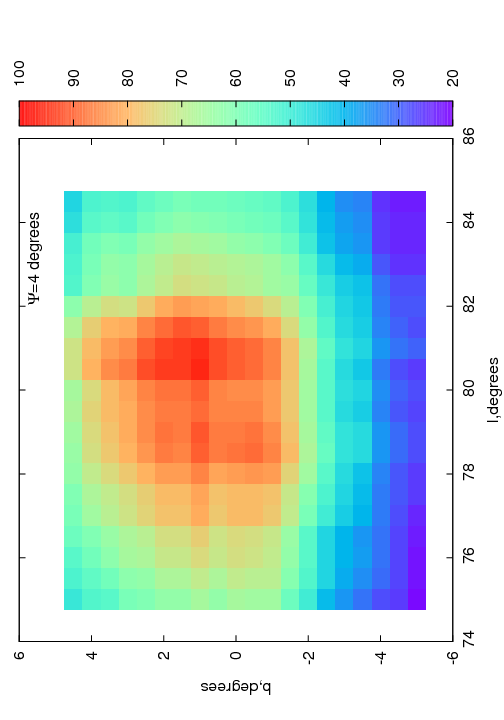}
\caption{Same as in Fig. \ref{fig:best_circle_2}, but for $\Psi=4^\circ$. }
\label{fig:best_circle_4}
\end{figure}

Since the bulk of the $\gamma$-ray emission from the Cygnus region is produced via cosmic ray interactions in the interstellar medium, not only the strength and the spectrum of the neutrino signal, but also its imaging characteristics could be derived directly from the $\gamma$-ray data. We have inspected the $\gamma$-ray data with the goal of localizing the circular "source" regions of the radii $\Psi=1^\circ, 2^\circ$ and $4^\circ$, which produce the highest gamma-ray (and hence neutrino) signal  above 100~GeV. Figs. \ref{fig:best_circle_2} and \ref{fig:best_circle_4} show the locations of the "highest signal" circles on the sky. Table \ref{tab:circles} lists the positions and expected neutrino fluxes of these circles.  The information of these two {\it a priori} coordinates of "highest signal" could be useful for IceCube in order to maximize the strength of the neutrino signal by correctly localizing the region used as the "source" region in the analysis of the data and limiting this way the trial factor associated with scanning the entire Cygnus region.

\begin{table}
\begin{tabular}{llllllll}
\hline
$\Psi$ & RA &DEC &l &b&$E_{thr}$& $A_{100}$&$\Gamma$\\
\hline
$2^\circ$ & 306.6$^\circ$ & 40.7$^\circ$ & 79.0$^\circ$  & 1.5$^\circ$  &$4.0\cdot 10^{4}$ &$1.05\cdot 10^{-12}$ & 2.40\\
$4^\circ$ & 308.3$^\circ$  & 41.7$^\circ$ & 80.5$^\circ$  & 1.0$^\circ$  &$3.2\cdot 10^{4}$& $1.65\cdot 10^{-12}$& 2.42\\
\hline
\end{tabular}
\caption{Optimal positions and energy threshold of the source signal regions for the search of neutrino emission from extended regions. The first column shows the radius of the source region circle, positions in equatorial and galactic coordinates are given in columns 2 to 4. The $6^{th}$ column shows the optimal energy threshold (in GeV), while the two last columns show the expected muonic neutrino signal deduced from the observed $\gamma$-ray signal (represented on the Fig.~\ref{fig:cygnus_spectrum}): normalization at 100GeV in units 1/GeV$cm^2$s and spectral index, respectively (see Eq.~(\ref{eq:spectrum})).}
\label{tab:circles}
\end{table}
\section{The detection of the neutrino signal in IceCube }


\subsection{IceCube sensitivity}

To find if the expected neutrino signal from these regions is detectable by IceCube, we compare the neutrino signal from each region with the atmospheric neutrino background for different energy bands $E>E_{thr}$, where $E_{thr}$ is an adjustable energy threshold for the IceCube data analysis. 

The value of $E_{thr}$ is fixed in such a way that it optimizes the signal to noise ratio. The spectrum of both the Cygnus region and of the low-declination region with high neutrino flux is harder than that of the atmospheric neutrino background. Therefore, the decrease of the atmospheric neutrino background with the increase of  $E_{thr}$ is faster than the decrease of the signal. At the same time, too strong increase of $E_{thr}$ reduces the signal to just a few counts, so that the detection efficiency of the source is strongly reduced. The optimal value of $E_{thr}$ is, therefore,  in between these two limits. Knowing the spectrum of neutrinos from any location along the Galactic Plane from the $\gamma$-ray data, we can adjust $E_{thr}$ as a function of Galactic longitude $l$ in such a way that the signal-to-noise ratio is maximized. 

Quantitatively, for each $E_{thr}$ and each $l$, we calculate the statistics of the source extended signal by convolving the known source spectrum given by Eq. (\ref{eq:spectrum}) with the energy- and declination-dependent area of IceCube, $A_{eff}(E,\delta)$, 
\begin{equation}
S(l,\Psi)=T\int_{E_{thr}}^\infty A_{eff}(E,\delta) \frac{dN_\nu(l,\Psi)}{dE}dE
\end{equation}
where $T$ is the exposure time and $\delta$ is the source declination, which is a function of the Galactic coordinates ($l$, $b$). To simplify the notation, we will only write here the Galactic longitude dependence, as we are focused on the Galactic Plane (($l$, $b=0^{\circ}$)$\equiv$($l$)).  In the same way we calculate also the background
\begin{equation}
\label{atmospheric}
B(\delta,\Psi)=\pi\psi^2 T\int^{\infty}_{E_{thr}} {A_{eff}(E,\delta)\frac{dN_{atm}(\delta,E)}{dEd\Omega}}dE,
\end{equation}
where $dN_{atm}(\delta,E)/(dEd\Omega)$ is the spectrum of the atmospheric neutrinos arriving from a declination $\delta$ in a solid angle $d\Omega$ \citep{honda2006}. 

We then calculate the Poisson probability to have at least $S(l,\Psi)+B(\delta,\Psi)$ neutrino counts from a given direction on the sky while the expected level of the atmospheric neutrino background is $B(\delta,\Psi)$. The "optimal" value of $E_{thr}(l)$ is chosen to minimize this probability for each Galactic longitude $l$, with an obvious additional requirement of having at least one signal count, $S(l,\Psi) \ge 1$.

\subsection{The neutrino signal from the Galactic Plane and from the Cygnus region}

\begin{figure}
\includegraphics[height=\columnwidth,angle=0]{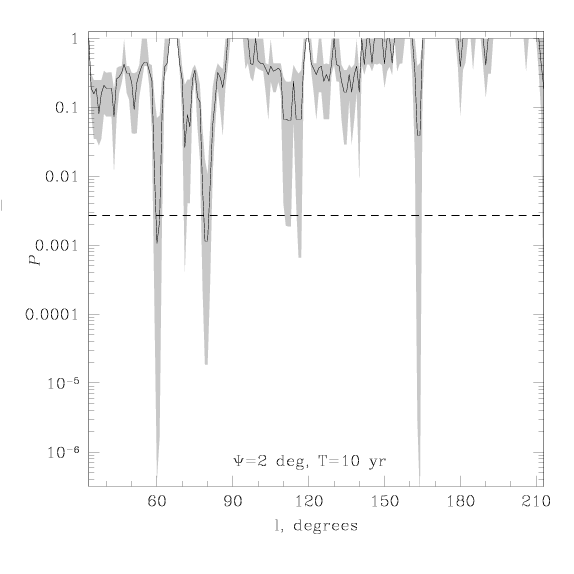}
\caption{Poissonian probability that the neutrino signal expected from the gamma-ray observations are due to the fluctuations of the atmospheric neutrino background for 10 years exposure time with IceCube in the IC-79 configuration and a 2 degree size region of interest around each longitude of the galactic plane ($l$), for the Northern hemisphere ($l\in[34:220]^\circ$). A line corresponding to the probability of 3$\sigma$ has been added to help the readability. Key: in black, the probability using the spectral index computed with the equation \ref{eq:Gamma}; in grey, the uncertainties of the probability corresponding to 1 sigma.} 
\label{fig:10yrs_2deg}
\end{figure}

\begin{figure}
\includegraphics[height=\columnwidth,angle=0]{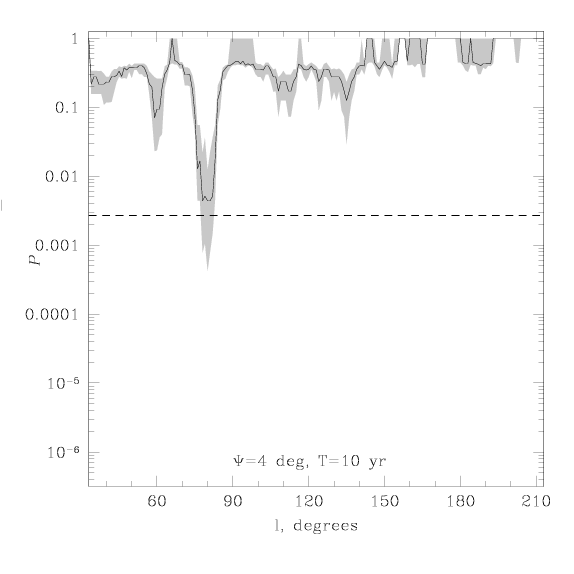}
\caption{Same as figure \ref{fig:10yrs_2deg}, but for a 4 degree size region of interest.}
\label{fig:10yrs_4deg}
\end{figure}

\begin{figure}
\includegraphics[height=\columnwidth,angle=0]{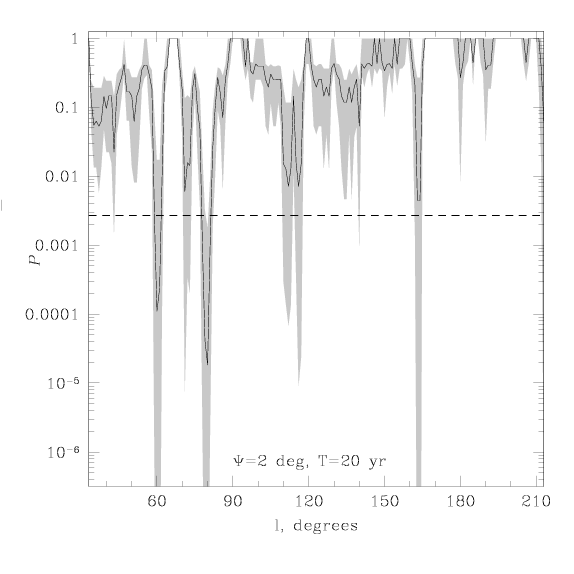}
\caption{Same as figure \ref{fig:10yrs_2deg}, but for 20 years exposure.}
\label{fig:20yrs_2deg}
\end{figure}

\begin{figure}
\includegraphics[height=\columnwidth,angle=0]{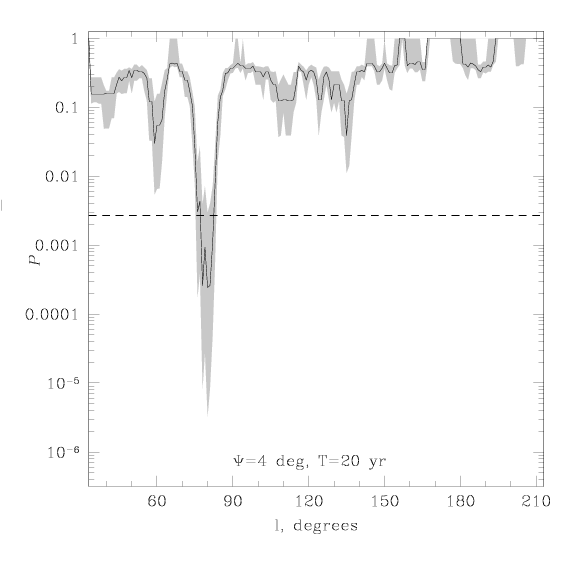}
\caption{Same as figure \ref{fig:20yrs_2deg}, but for a 4 degrees size region of interest.}
\label{fig:20yrs_4deg}
\end{figure}

The results of such optimization procedure for different exposure times of 10 and 20 years and different choices of the source region circle radius $\Psi$ are shown on the Figs~\ref{fig:10yrs_2deg}, \ref{fig:10yrs_4deg}, \ref{fig:20yrs_2deg} and \ref{fig:20yrs_4deg}. In this figures we plot the Poisson probability which measures the inconsistency of the source+background hypothesis ($S(l,\Psi)+B(\delta,\Psi)$) with the background-only hypothesis.  The calculations are done for the 79 string (IC-79) configuration of IceCube detector for which the effective areas were published \citep{effareaIC79}. The IC-79 effective areas are close to those of the final configuration with 86 strings (IC-86), so that the potential of IceCube for the detection of the neutrino emission from the Galactic Plane could be correctly estimated based already on the known IC-79 instrument characteristics.
 

From Figs. \ref{fig:10yrs_2deg} and \ref{fig:10yrs_4deg}  one could see that, as expected, the strongest excess (i.e. the strongest inconsistency with the background-only hypothesis) is found in the Cygnus region. A ten-year exposure is, however, not sufficient for a detection of the Cygnus region at the $3\sigma$ level, except if the spectral index remains as hard as the one detected with Fermi at higher energies. As shown on the figure~\ref{fig:10yrs_4deg}, somewhat surprisingly, the low-declination region at $l=33^\circ$ with a stronger flux in the 100~GeV band (see Fig. \ref{fig:profile_gamma}) gives a weaker excess.  This is explained by a softer spectrum of the source, which results in a lower number of source counts in the IceCube energy range.  

Figs. \ref{fig:20yrs_2deg} and \ref{fig:20yrs_4deg} show that a twenty-year exposure of IceCube should be sufficient for a detection of the Cygnus region at $3\sigma$ level, for both source regions of radius $\Psi=2^\circ$ and  $\Psi=4^\circ$(marginally), centered on the position given in Table \ref{tab:circles}. The probability of detection is higher if the source region is smaller, this is explained by a higher level of atmospheric neutrino background than signal events in this larger region. 


It is important to note that apart from the Cygnus, none of the other regions of the Galactic Plane will be detectable even with a twenty year exposure of IceCube. It appears that the part of the Galactic Plane in the Northern hemisphere is not a bright neutrino source. Since the estimate of neutrino spectrum based on the observed $\gamma$-ray spectrum is relatively precise, there is practically no uncertainty in this conclusion. 

\subsection{The neutrino signal from the inner Galaxy with a hypothetical detector in the Northern Hemisphere}

The level of neutrino emission from the part of the Galactic Plane situated in the Northern Hemisphere is so low because this part of the Galaxy is mostly in the "outer Galaxy" direction. Only the Cygnus region is in the direction of the nearby Galactic arm, which increases the column density of the cosmic rays and interstellar medium in this direction. The strongest $\gamma$-ray and neutrino signal come from the inner part of the Galactic Plane which is entirely situated in the Southern Hemisphere.  

Unfortunately, the IceCube detector is optimized for the sources located in the Northern Hemisphere (see effective area in the paper of \citep{effareaIC79} so that the brighter part of the Galactic Plane is not accessible. The main obstacle for the observations at negative declinations is the strong residual atmospheric muon background which strongly dominates at low energies and effectively increases the energy threshold of IceCube to the PeV region. The spectrum of the Galactic diffuse neutrino emission is relatively soft, so that most of the signal statistics comes at the low energy threshold.

To demonstrate the potential of neutrino astronomy which might be done with a km$^3$ class neutrino detector in the Northern hemisphere,  we performed the analysis identical to that reported in the previous subsection, but assuming a hypothetical "IceCube-like"  neutrino detector situated at the North rather than the South Pole. One candidate for such km$^3$ class neutrino detector in the Northern Hemisphere is KM3NeT project to be located in the Mediterranean Sea and which is currently in the design phase \citep{km3net}. 

\begin{figure}
\includegraphics[height=\columnwidth,angle=0]{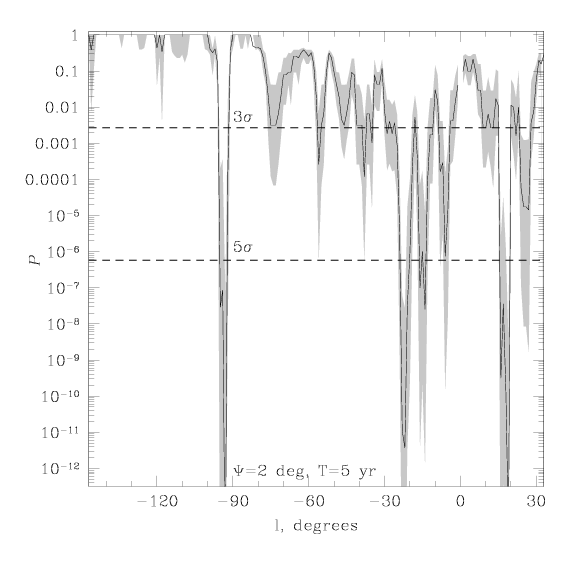}
\caption{Same as in figure \ref{fig:10yrs_2deg}, but for a hypothetical "Northern IceCube" detector situated at the North Pole and sensitive for the sources in the Southern hemisphere. For 5 years of exposure.}
\label{fig:5yrs_2deg_south}
\end{figure}

\begin{figure}
\includegraphics[height=\columnwidth,angle=0]{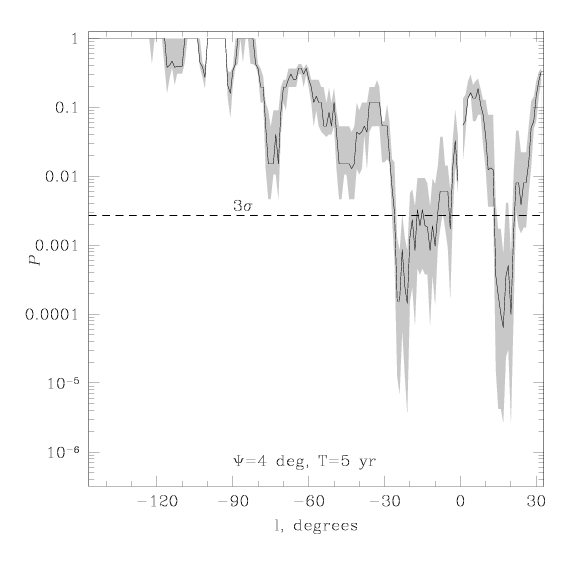}
\caption{Same as in figure \ref{fig:5yrs_2deg_south}, but for a 4 degree size region of interest around each galactic longitude.}
\label{fig:5yrs_4deg_south}
\end{figure}

The results of the calculations for the "Northern IceCube" are shown in Figs. ~\ref{fig:5yrs_2deg_south} and \ref{fig:5yrs_4deg_south} for five years exposure. One could see a dramatic difference with the expected IceCube results. A detector identical to IceCube, but in the Northern hemisphere would detect some $\sim 3$ sources at the $5\sigma$ level already with just five years of operation (compared with the ten-to-twenty years needed for a 3$\sigma$ evidence of single source in real IceCube), if the region of interest is of 2 degrees. For some of the detected sources, the imaging analysis might be possible with a five-year data set: one could see that several sources would be significantly detectable in both $\Psi=2^\circ$ and $\Psi=4^\circ$ regions.

This implies that a limited imaging analysis of the neutrino signal from Southern hemisphere will be possible with a hypothetical "IceCube-like"  neutrino detector situated at the North Pole. One will be able to compare the morphology of neutrino and $\gamma$-ray signals from this region. 

It is interesting to note that the Galactic Center itself would not be detected by the "Northern IceCube". This is explained by the relatively soft spectrum of the source in the Galactic Center, see Fig. \ref{fig:spectrum_gamma}. In general, all the isolated extended sources detectable at 5$\sigma$ level within a five-year exposure by the "Northern IceCube" are locations with a hard spectrum of emission. 

It is clear that detection of numerous VHE neutrino sources in the inner Galactic Plane would provide rich astronomical data on the locations of high-energy particle accelerators of protons and atomic nuclei in the "cosmic scale" particle accelerators in the Galaxy and a true start of the field of "multi-messenger" astronomy. 

The table \ref{table:south} summarizes the position and neutrino spectral information of the sources that could be detectable by a neutrino telescope located in the Northern Hemisphere such as KM3NeT.
In the next section will be described those potential neutrino sources.

\begin{table}
\small
\begin{tabular}{llccc}
\hline
Gal. longitude ($l$)& $Name$ & $E_{thr}$& $A_{100}$&$\Gamma$\\
19$^{\circ}$&HESS J1825-137& 2.5$\cdot 10^{4}$&3.0$\cdot 10^{-12}$& 2.30\\
 267$^{\circ}$&Vela Jr.& 3.2$\cdot 10^{4}$&1.1$\cdot 10^{-12}$& 2.12\\
 337-338$^{\circ}$&HESS cluster & 2.0$\cdot 10^{4}$&$\sim$3.2$\cdot 10^{-12}$& 2.33\\
\hline

\hline
\end{tabular}
\caption{List of the sources located in the Southern hemisphere potentially detectable with an detector similar to the IceCube detector located in the Norther Pole (see Fig.~\ref{fig:5yrs_2deg_south}). The three last columns show the expected neutrino flux (see Eq.~(\ref{eq:spectrum})) and the optimal energy threshold for a considered region of 2 degrees around the Galactic longitude listed in the first column. Units for $E_{thr}$ and $A_{100}$ are the same than in table \ref{tab:circles}. The 'HESS cluster' is a generic term for the sources detected by HESS in the region around $l=337-338^{\circ}$.} 
\label{table:south}
\end{table}

\subsection{Neutrino candidate sources located in the Southern hemisphere:}
Sources listed in table~\ref{table:south} could be sites of recent cosmic rays acceleration. 
Among them, the HESS J1825-137 is the brightest gamma-ray source (see Fig.~\ref{fig:profile_gamma}) but with a soft spectral index. Vela Jr., on the other hand, is expected to have the hardest spectrum which will make it detectable at the same level even with a lower flux. 
In the following, we will describe each of these neutrino candidate sources.

The source \textit{HESS J1825-137} has been detected with the H.E.S.S. telescope \citep{HESS_survey} as an extended source. This source could be associated with an X-ray pulsar wind nebula G18.0--0.7 \citep{HESS_J1825PWN}.  In \citet{HESS_ J1825pulsar}, the authors associate the presence of those two sources with the Vela-like pulsar PSR B1823-13 and explain this emission as a result of a SNR expanding and evolving into an inhomogeneous medium. 

The supernova remnant \textit{'Vela Jr.'}, also know under the name RX J0852.0-4622, is situated at the Galactic coordinate ($l=267^{\circ}$, $b=-01.22^{\circ}$). It is one of the most luminous galactic sources in the VHE energy band (the flux was reported to be of about $10\%$ of Crab at 1 TeV by \citet{VelaJr_Cangaroo}). It has been discovered in 1998  using X-ray images performed by ROSAT \citep{VelaJr_dec}. The distance to Vela Jr. has been estimated to be of the order of 200 pc \citep{VelaJr_param}, which is closer than the Cygnus region. The CANGAROO-II imaging atmospheric Cerenkov telescope detected the source at 6 $\sigma$ level in 100 hours \citep{VelaJr_Cangaroo}. This source has been also detected by the H.E.S.S. telescope \citep{VelaJr_hess,VelaJr_proceedings} with a differential gamma-ray spectrum which follows a power law distribution of a spectral index $\sim$2.1. This flux measurement is consistent with the flux obtained at lower energy with the Fermi telescope (see table \ref{table:south}). According to \citet{VelaJr_param}, this remnant is quite young, with an age of 680 years, and thus could be the site of recently accelerated cosmic rays.

The extended emission detected by the Fermi telescope at \textit{Galactic longitude around $l=337-338$} has a counterpart at higher energy.
The inner Galactic plane survey of H.E.S.S. \citep{hess_survey06} revealed several sources located in the considered region: HESS J1632-478 ($l=336.38^{\circ}$, $b= 0.19^{\circ}$), HESS J1634-472 ($l=337.11^{\circ}$, $b=0.22^{\circ}$)  and HESS J1640-465 ($l=338.32^{\circ}$, $b=Ð0.02^{\circ}$).

\begin{itemize}
\item{}The pulsar PSR J1632-4818  has been detected close to the position ($l=336.08^{\circ}$, $b=-0.21^{\circ}$), with an age estimated to be of the order of  20 kyr \citep{PSRsurvey}. Another close pulsar was detected, PSR  J1632-4757, with a spin-down loss of $5\cdot10^{34}$ ergs/s and a distance of 7 kpc.
Both pulsars could be marginally associated to the source HESS J1632-478  \citep{hess_survey06}. It seems thus possible that the emission from this region is composed of contributions from many sources.

\item{}The source HESS J1634-472 has been detected during this same survey \citep{hess_survey06} as an extended gamma-ray source with a flux of 6$\%$ of the Crab Nebula and with a position that agrees with the supernova remnant G337.2+0.1.

\item{}The gamma-ray source HESS J1640-465 has also been observed in the radio band. Using the radio observations at 843 MHz \citep{HESS1640_radio}, the source position coincides roughly with the supernova remnant G338.3-0.0. This low energy counterpart to the VHE emission could be due to a supernova explosion followed by a pulsar formation.
\end{itemize}

Those sources are potentially the sites of recent injection of cosmic rays and thus, they should be good candidates for the production of neutrinos. According to our analysis, if it is the case, those sources should be detectable with a neutrino detector having similar performance as the IceCube detector, but sensitive to sources located in the Southern hemisphere (see table \ref{table:south}).

\section{Conclusions}

We have investigated the detectability of the neutrino diffuse emission produced by the cosmic ray interactions in the Galaxy. Quantitative calculation of the spectrum of the neutrino signal from each direction along the Galactic Plane were done based on the Fermi $\gamma$-ray data in the 100~GeV band. We find that a possible evidence for neutrino detection from the Cygnus region can be reached only after about 20 years of exposure. Our estimates show also that the diffuse neutrino emission form the outer part of the Galaxy visible to IceCube is too weak to be detected as a combination of unidentified neutrino sources within a generic point-source analysis technique.  Nevertheless for the Cygnus region we have found precise positions, imaging and spectral characteristics (table \ref{tab:circles}), which could be used to optimize the IceCube data analysis to facilitate the source detections. Unlike for sources located in the Northern hemisphere, the detection of neutrino sources located in the Southern hemisphere seems more favorable. In our analysis, we found several neutrino sources which should be detectable with a neutrino detector sensitive to sources located in the Southern hemisphere within only a few years of exposure (table \ref{table:south}).

Indeed, our analysis of neutrino signal from the Galaxy detectable with an IceCube-like detector in the Northern hemisphere shows that rich astrophysical observational program would be available for a such detector.  Some $\sim 3$ sources would be detectable in $\sim 5$~yr exposure time, with the potential for the detailed imaging / spectral studies, which would lead to a significant progress toward identification of sources of Galactic cosmic rays.

It should be mentioned that the analysis presented here is based on a binned method both in space and energy and it assumed the 79-string effective areas of IceCube. However, IceCube analysis of their data is normally perform using a more sophisticated unbinned method, where the energy and spatial distribution of events is used in the likelihood maximization procedure to enhance the discovery potential. 


\end{document}